\documentclass[aps, twocolumn, showpacs, amsmath]{revtex4}
\usepackage{dcolumn}
\usepackage{bm}
\usepackage{graphicx}
\usepackage{color}
\usepackage{subfigure}
\usepackage{hyperref}
\usepackage{latexsym}
\usepackage{amsthm}
\usepackage{amssymb}

\DeclareGraphicsExtensions{.jpg,.pdf, .mps, .png, .eps, .ps, .EPS,.gif}

\begin{document}
\def\be{\begin{equation}}
\def\ee{\end{equation}}

\def\bc{\begin{center}}
\def\ec{\end{center}}
\def\bea{\begin{eqnarray}}
\def\eea{\end{eqnarray}}
\newcommand{\avg}[1]{\langle{#1}\rangle}
\newcommand{\Avg}[1]{\left\langle{#1}\right\rangle}

\def\ie{\textit{i.e.}}
\def\etal{\textit{et al.}}
\def\m{\vec{m}}
\def\G{\mathcal{G}}

\title{   Percolation in Multiplex  Networks with Overlap}

\author{Davide Cellai$^1$, Eduardo L\'opez$^{2,3}$, Jie Zhou$^1$, James P.~Gleeson$^1$ and  Ginestra Bianconi$^4$}

\affiliation{$^1$MACSI, Department of Mathematics and Statistics, University of Limerick, Ireland\\
 $^2$CABDyN Complexity Centre, Sa\"id Business School, University of Oxford, Oxford OX1 1HP, UK \\
$^3$Physics Department, Claredon Laboratory, University of Oxford,  Oxford OX1 3PU, UK\\
$^4$School of Mathematical Sciences, Queen Mary University of London, London, E1 4NS, UK}
\begin{abstract}
From  transportation networks to complex infrastructures, and to social and communication networks, a large variety of systems can be described in terms of  multiplexes formed by a set of nodes interacting through different networks (layers). Multiplexes may display an increased fragility with respect to the  single layers that constitute them.
However, so far the overlap of the links in different layers has been mostly neglected, despite the fact that it is an ubiquitous phenomenon in most multiplexes.
Here we show that the overlap among layers can improve the robustness of interdependent multiplex systems and  change the critical behavior of the percolation phase transition in a complex way.
\end{abstract}
\pacs{89.75.Fb, 64.60.aq, 05.70.Fh, 64.60.ah}
\maketitle
\section{Introduction}
In the last decade we have gained a deep understanding of the interplay
between the topology of single complex networks \cite{RMP,Newman_rev,Boccaletti2006}  and the behavior of critical phenomena occurring on them  \cite{crit,Dynamics}.
Recently,  it has become clear
that in order to fully investigate the  properties  of a large variety
of complex systems such as energy supply networks \cite{Havlin1}, complex infrastructures \cite{Rinaldi,rosato2008,Kurant,morris2012,Boccaletti}, social networks \cite{Thurner}, climatic systems \cite{Kurths} and
brain networks \cite{Bullmore2009}, it is necessary to consider their multilayer structure \cite{Mucha}.
Each layer corresponds to a network with a specific function, but the entire
complex system requires multiple layers operating in a coupled way and
forming an interacting set of networks.
For example, analyses of the disruptions provoked by an earthquake to large infrastructures show the relevance of interdependence among power transmission and telecommunications \cite{duenas2012} and the different resilience of coupled networks such as the power grid and the water system \cite{hernandez2011}.

An important and ubiquitous example of multilayer networks is a multiplex where the same nodes are linked by different networks (layers).
A multiplex is formed by a set of $N$ nodes and $M$ layers. Every node
is represented in every layer, and every layer is formed by a network
of interactions (links) between the nodes. For instance, multiplexes are good
descriptions of social networks, where the nodes represent agents and the different
layers correspond either to different types of interaction (such as
family, work or friendship ties)\cite{Thurner}
or to different means of communication (email, chat, mobile phone, etc.\dots) \cite{MultiplexPR}.
Other examples of multiplexes can be found in transportation networks,
where cities can be connected by roads, railways, waterways or airline connections \cite{morris2012,Boccaletti,woolley2011},
or brain networks where different regions of the brain belong at the same time
to the functional network of brain activity and to the structural brain network \cite{Bullmore2009}.

In the last years, interest on multiplexes has been growing both on the theoretical side, and on the empirical side of multiplex data analysis. In fact different models of multiplex structure  have been formulated \cite{Growth1,Growth2,PRE,dedomenico2013},
critical phenomena and dynamical processes have been characterized on multiplex  structures \cite{Havlin1,Havlin2,Havlin2b,Havlin3,parshani2010,baxter2011,cellai2013a,bollobas2001,baxter2012,Son,Kabashima,Leicht,JSTAT,Diffusion,funk2010,Boguna,Cooperation,cozzo2013},
and finally new structural measures have been introduced and evaluated in large multiplex datasets \cite{Boccaletti,morris2012,Thurner,Kurths}.
The theoretical interest has been partly sparked by the discovery \cite{Havlin1,Havlin2,Havlin2b,Havlin3} that
the robustness properties of
a multiplex formed by interdependent networks is strongly affected by its
multilayer structure. In fact a multiplex can be much more fragile than single
networks and its functionality can be affected significantly by cascades of node failures \cite{Havlin1,Havlin2,Havlin3}.
In this context, the mutually connected giant component (MCGC) plays a pivotal role.
The MCGC of a multiplex is the extensive component where every pair of nodes
is connected in every layer by at least one path formed by nodes inside the MCGC.
If a fraction $1-p$ of nodes is removed from a multiplex with a MCGC, the
size of the MCGC is reduced.
At some value of $p$, a critical point is reached where a  first
order hybrid transition is observed and the MCGC abruptly drops from a finite
value to zero \cite{Havlin1,Havlin2,Havlin2b,Havlin3,baxter2012,Son,Kabashima}.

Despite the interesting theoretical behavior, analysis of the empirical side
offers another observation not yet addressed by the theory, which is
that many multiplexes \cite{Thurner,Boccaletti,parshani2010} are formed by correlated networks characterized by
a significant overlap of the links. For example, in social networks  it is common for
two friends to communicate both via email and via mobile phone, or in transportation
networks two cities connected by a main road are likely to be connected
also by a railway.
Given such empirical findings, the theory of multiplex networks cannot be complete until
the effects of link overlap in the multiplex robustness properties (characterized by the MCGC)
are understood.

Our aim is not to address a specific application, but give instead a general theoretical approach that can include edge overlap in the study of percolation under random damage.
Percolation is perhaps the simplest model of robustness and stability, but it constitutes a first step in dealing with more complex models and even dynamical processes occurring on the network.
Therefore, introducing a formalism with edge overlap in percolation on multiplex networks should be seen as an important building block of the science of complex networks.

In this paper, we apply this formalism to the case of 2 and 3-layer Poisson graphs and calculate the phase diagrams of the models.
We show that edge overlap can significantly enhance  the robustness properties of the multiplex.
We also show that a multiplex formed by more than two layers presents a very rich phase diagram with
high order critical points, characterizing the increased complexity of this percolation problem.

In Section II we introduce the notation of multiplexes with edge overlap, in Section III we formulate a general framework to characterize the emergence of the MCGC, in Section IV we calculate the phase diagrams of percolation on 2 and 3-layer Poisson graphs and in Section V we draw the conclusion of this work.

\section{  Multiplex with Overlap}
Consider a multiplex formed by $N$ labelled nodes $i=1,2\ldots, N$
and $M$ layers.
We can represent the multiplex as described in \cite{Mucha}.
To this end we indicate by $\vec{\G}=(\G^1,\ldots, \G^{\alpha},\ldots, \G^M)$ the set
of all the networks $\G^{\alpha}$ at layer $\alpha=1,2,\ldots, M$ forming
the multiplex. Each of these networks has an adjacency matrix
with matrix elements $a_{ij}^{\alpha}=1$ if there is a link between node
$i$ and node $j$ in layer $\alpha$ and zero otherwise.

\begin{figure}
\begin{center}
\includegraphics[width=0.8\columnwidth]{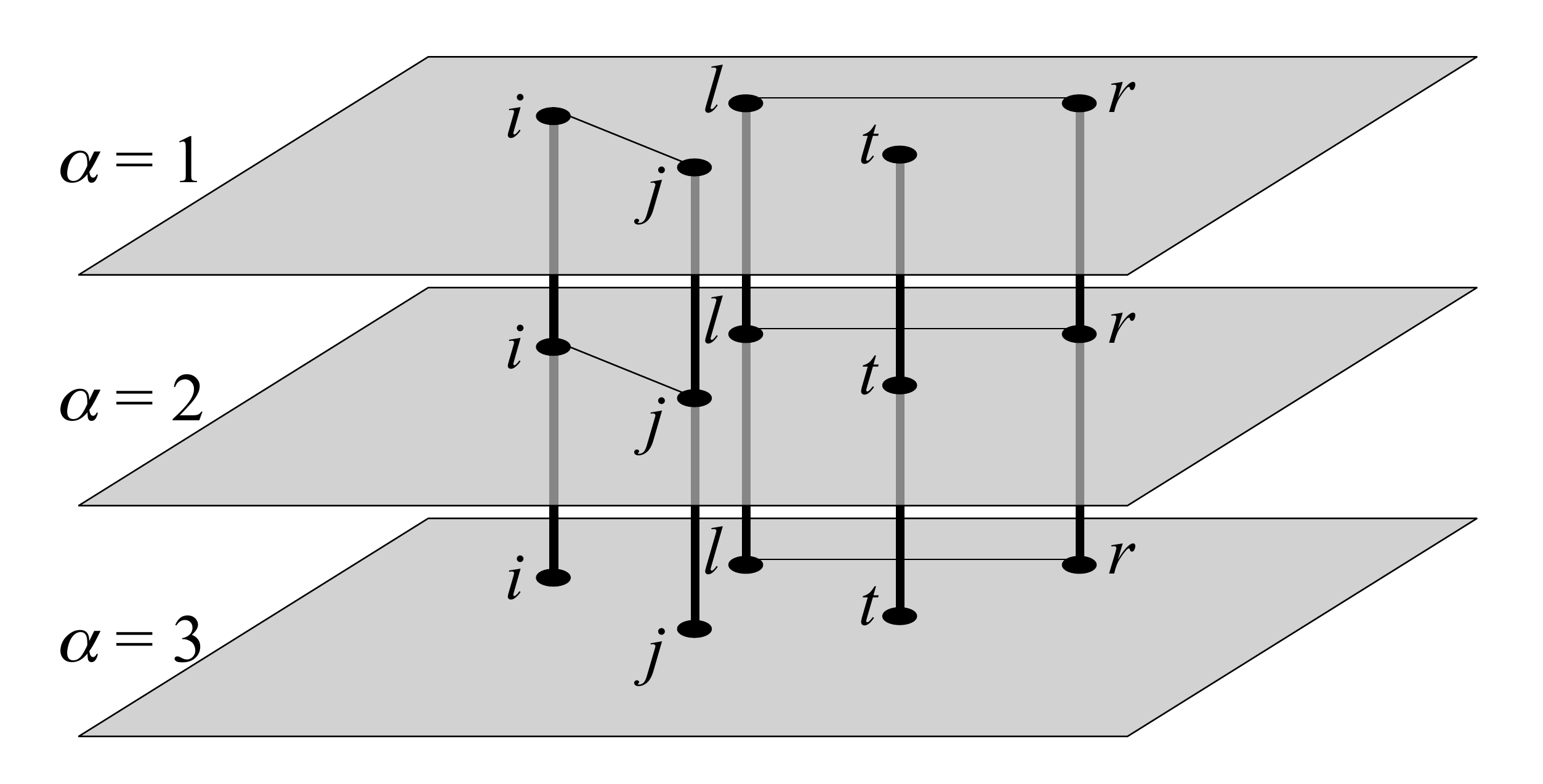}
	\caption{In this example of multiplex network with three layers and  overlap of the links the nodes $(i,j)$ are linked by a multilink $\vec{m}=(1,1,0)$, the nodes $(r,l)$ are linked by a multilink $\vec{m}=(1,1,1)$, all the other pairs of nodes are linked by a multilink $\vec{m}=(0,0,0)$. }
	\label{fig:1}
\end{center}
\end{figure}

A useful concept to characterize multiplex structure are {\it multilinks},
and {\it multidegrees} recently defined in \cite{PRE}.
Let us consider the vector $\vec{m}=(m_1,\ldots,m_{\alpha},\ldots, m_M)$
in which every element $m_{\alpha}$ can take only two values $m_{\alpha}=0,1$.
We define a {\it multilink}, and we represent it with $\vec{m}$, as the set of links connecting a
given pair of nodes in the different layers of the multiplex and connecting them in the  generic layer $\alpha$ only if $m_{\alpha}=1$.
Multilinks are mutually exclusive, i.e.  any pair of nodes $(i,j)$ can be linked only by one multilink $\vec{m}$, that we call $\vec{m}^{ij}$.
In Figure $\ref{fig:1}$ we show an example of a multiplex where nodes $(i,j)$ are linked by a multilink $(1,1,0)$ and nodes $(r,l)$ are linked by a multilink $(1,1,1)$.

Note how the introduction of the multilink now makes explicit the notion of overlap of the links:
the links $(i,j)$  in the layers $\alpha_1,\dots,\alpha_t$ overlap if  the nodes $i$ and $j$ are joined by a multilink $\vec{m}=\vec{m}^{ij}$ and $m_{\alpha_1}^{ij}=\dots=m_{\alpha_t}^{ij}=1$.

For a multiplex it is possible to define a set of multiadjacency matrices of elements { by setting} $A_{ij}^{\vec{m}}=1$ if node $i$ is linked to node $j$ by a multilink $\vec{m}$ and $A_{ij}^{\vec{m}}=0$ otherwise.
In other words, every element $A_{ij}^{\vec{m}}$ not only depends on the node pair $(i,j)$ as in classical adjacency matrices, but it is also a function of the generic multilink $\m$.
These multiadjacency matrices  can be defined in terms { of} the adjacency matrices $(a_{ij}^{\alpha})$ on each layer $\alpha$, {\ie} for each $\vec{m}=\{0,1\}^M$,
\bea
	A_{ij}^{\vec{m}}=\prod_{\alpha=1}^M \left[a_{ij}^{\alpha}m_{\alpha}+(1-a_{ij}^{\alpha})(1-m_{\alpha})\right].
\eea
The expression in the square brackets (in  an unweighted network) can only be zero or one.
It equals one only when the components of $\m$ match the actual presence of an $(i,j)$ link on the layer $\alpha$.
Otherwise, the expression is zero and correctly represents the absence of a multilink of given type $\m$ between $i$ and $j$.
The number of multilinks is $2^M$ but the elements of the multiadjacency matrices are not all independent.
In fact they satisfy the relation
\bea
\sum_{\vec{m}\in \{0,1\}^M}A_{ij}^{\vec{m}}=1,
\label{norm}
\eea
 where the sum is over the set of all possible multilinks. Since the elements of the multiadjacency matrices $A_{ij}^{\vec{m}}$ are either zero or one, 
 the condition given by Eq.$(\ref{norm})$ implies that between any pair of nodes $(i,j)$ there is a single multilink that we indicate with $\vec{m}^{ij}$.
Therefore only $2^M-1$ adjacency matrices are independent.
In most multiplex networks the number of layers is finite, therefore the exponential number of multilinks is not an important limitation to this approach. Moreover, our framework can be easily generalized to the case  of multiplexes with a large number of layers $M$ in which only few types of multilinks are allowed.

Finally, we define the {\it multidegree} with respect to $\vec{m}$
of a node $i$, $k_i^{\vec{m}}$, as the total number of multilinks $\vec{m}$
{ connected to} node $i$, i.e.
\bea k_i^{\vec{m}}=\sum_{j=1}^N A_{ij}^{\vec{m}}.
\eea
In the specific case of a duplex ($M=2$) we  have
$ k_{i}^{(1,1)}=o_i$, where $o_i=\sum_j a^1_{ij}a^2_{ij}$ is the local overlap of node $i$, i.e. the total number of nodes $j$ linked to node $i$ both in layer 1 and in layer 2.
In general, for a multiplex of $M$ layers the multidegree $k^{\vec{m}}$ can be considered
as a high order measure of local overlap as long as $\sum_{\alpha}m_{\alpha}>0$.

This framework encapsulates two particular cases, which are already known to the scientific community.
First, let us consider the case of fully overlapping multiplexes, where all links are the same in all layers.
In this case, we recover classical percolation with a continuous second order phase transition.
Second, we may consider the case where edges on different layers are totally uncorrelated.
As we are considering configuration model graphs on each layer, the multiplex consists of a random coupling of locally tree-like, sparse random graphs \cite{bollobas2001}.
Therefore, the probability of edge overlap vanishes as $N\to\infty$.
It has been recently shown that in this case the phase transition is discontinuous \cite{baxter2012}.
In general, instead, one needs to consider overlap effects when designing or studying the fragility of multiplexes.
Our formalism, and in particular the concept of multilink, can be very helpful, giving us a fully controllable tool to study the problem.

\section{ Emergence of the Mutually Connected Giant Component (MCGC)}

As we mention in the introduction, the MCGC of a multiplex is the extensive component where every pair of nodes
is connected in every layer by at least one path formed by nodes inside the MCGC.
Then, we can consider the behavior of the size of the MCGC in a multiplex as a fraction $1-p$ of nodes is removed.
If there is no link overlap, at a critical value of $p$, a first
order hybrid transition is observed and the MCGC abruptly drops from a finite
value to zero \cite{Havlin1,Havlin2,Havlin2b,Havlin3,baxter2012,Son}.

Our goal is to characterize the effect of link overlap on how the mutually connected giant component (MCGC) changes as a function of $p$ (the fraction of remaining nodes).
Specifically, we will follow and extend the approach developed  by Son {\etal} \cite{Son} for studying the percolation on two interdependent networks without overlap.
First of all, we observe that a node $i$ belongs to the mutually connected giant component of the multiplex  networks if and only if the { following} $M$  conditions are met simultaneously: for every layer $\alpha=1,2,\ldots, M$
{\it  at least  one of the neighbor nodes of $i$ in network $\alpha$ belongs to  the mutually connected giant component of the interdependent networks}.
We consider a tree-like multiplex, {\ie} a multiplex in which the { combined} network of all the layers ({\ie} the network of adjacency matrix $B_{ij}=\theta(\sum_{\alpha=1,M} a_{ij}^{\alpha})$,  where $\theta(x)=1$ if $x>0$ otherwise $\theta(x)=0$)  is locally tree-like.
According to the above recursive definition of the MCGC,  on a locally tree-like multiplex we can determine if a node belongs to the MCGC by a message-passing algorithm \cite{Mezard,Kabashima}, described in Son {\etal} as an ``epidemic spreading'' process \cite{Son}.

Let $s_i=0,1$ indicate if a node is removed or not from the network and let   $\sigma_i=0,1$ be the indicator function that the node $i$ is in the mutually connected giant component. The value of $\sigma_i$ is determined by the ``messages" that the neighboring nodes send to node $i$. We denote these ``messages" as $\sigma_{j\to i}^{\vec{m}^{ij}}$. The value of the message is set to one $\sigma_{j\to i}^{\vec{m}^{ij}}=1$ if and only if the following two conditions are satisfied:
\begin{itemize}
	\item[(a)] node $j$ is a neighbor of node $i$ with a multilink $\vec{m}^{ij}$ connecting them such that $\sum_{\alpha}m_{\alpha}^{ij}>0$;
	\item[(b)] node $j$ belongs to the mutually connected giant component even if the multilink $\vec{m}^{ij}$ between node $j$ and node $i$ is removed from the multiplex.
\end{itemize}
Otherwise we will have $\sigma_{j\to i}^{\vec{m}^{ij}}=0$.

In a locally  tree-like multiplex the value of $\sigma_i$ and $\sigma_{j\to i}^{\vec{m}^{ij}}$ satisfy the following message-passing relations
\begin{equation}
	\sigma_i = s_i \prod_{\alpha=1}^M\left[1-\prod_{j\in N_{\alpha}(i)}(1-\sigma_{j\to i}^{\vec{m}^{ij}})\right],
	\label{sigma-1}
\end{equation}
\begin{equation}
	\sigma_{j\to i}^{\vec{m}^{ij}} = s_j \prod_{\alpha=1}^M\left[1-\prod_{l\in N_{\alpha}(j) \setminus i}(1-\sigma_{l \to j}^{\vec{m}^{l j}})\right],
	\label{sigma-ijm-1}
\end{equation}
where $N_{\alpha}(i)$ indicates the set of nodes that are neighbor of node $i$ in layer $\alpha$.
The expression within square brackets in Eq. (\ref{sigma-1}) is one only if   on layer $\alpha$ at least one of the neighbors $j$ of node $i$ is connected to the MCGC { by an edge other than} the $(i,j)$ edge, otherwise is zero.
Equation (\ref{sigma-ijm-1}) has been written using the same logic, but taking now into consideration that $j$ is required to be in the MCGC by { connecting to a } neighbor $l$ which must be different from $i$.
This is calculated by subtracting from one the probability that all the neighbors $j$ of $i$ are not attached to the MCGC.

Let us now consider equation (\ref{sigma-ijm-1}), by using the formula
\begin{equation*}
	\prod_{\alpha=1}^M (1-x_i) = \sum_{\vec{r}}(-1)^{\sum_{\alpha}r_{\alpha}} x_1^{r_1}\dots x_M^{r_M},
\end{equation*}
where $\vec{r}=(r_1,r_2,\ldots, r_{\alpha},\ldots r_M)$ with $r_{\alpha}=0,1$,
and where the sum $\sum_{\vec{r}}$ is over all possible vectors $\vec{r}${. We} can expand the multiplications and write
\begin{equation}
	\sigma_{j\to i}^{\vec{m}^{ij}} = s_j \sum_{\vec{r}}(-1)^{\sum_{\alpha}r_{\alpha}}\prod_{\alpha=1}^M\prod_{l\in N_{\alpha}(j) \setminus i}(1-\sigma_{l \to j}^{\vec{m}^{lj}})^{r_{\alpha}}.
	\label{sigma-ijm-2}
\end{equation}
By observing that for each node $l$ {neighboring} node $j$ in layer $\alpha$ we should necessarily have ${m}^{lj}_{\alpha}=1$ and that $\sigma_{l\to j}^{\vec{m}^{lj}}=0,1$, it is possible to swap the two products in equation (\ref{sigma-ijm-2}) and show that
\begin{equation}
	\sigma_{j\to i}^{\vec{m}^{ij}} = s_j\sum_{\vec{r}}(-1)^{\sum_{\alpha}r_{\alpha}}\prod_{
		\begin{subarray}{c}
        			l=1\\
			l\neq i
      		\end{subarray}}^N (1-\sigma_{l \to j}^{\vec{m}^{lj}})^{\sum_{\alpha}r_{\alpha}m^{lj}_{\alpha}}.
      		\label{sij}
\end{equation}
Similarly, we can show that
\begin{eqnarray}
	\hspace*{-5mm}\sigma_{i} &=& s_i\sum_{\vec{r}}(-1)^{\sum_{\alpha}r_{\alpha}}\prod_{l=1}^N(1-\sigma_{l \to i}^{\vec{m}^{li}})^{\sum_{\alpha}r_{\alpha}m^{li}_{\alpha}}.
	\label{s2}
\end{eqnarray}

Let us now consider a   multiplex belonging to the multiplex ensemble \cite{PRE} characterized by a given multidegree distribution and no degree-degree (multidegree-multidegree) correlations. We can construct a network in this ensemble by extending the configuration model to networks with multidegrees.
To this end, we draw the multidegree sequence $\{k_i^{\vec{m}}\}$ from a multidegree distribution $P(\{k^{\vec{m}}\})$. We then  attach $k_i^{\vec{m}}$ stubs of type $\vec{m}$ such that  $\sum_{\alpha}m_{\alpha}>0$ to each node $i$. Finally  we randomly match the stubs of the same type of multilinks belonging to  different nodes.
In this ensemble the probability $P(\vec{\G})$ of a multiplex $\vec{\G}$ is given by
\bea
P(\vec{\G})=\prod_{ij}\prod_{\vec{m}\neq \vec{0}}\left(\frac{k_i^{\vec{m}}k_j^{\vec{m}}}{\avg{k^{\vec{m}}}N}\right)^{A_{ij}^{\vec{m}}}\left[1-\sum_{\vec{m}\neq \vec{0}} \frac{k_i^{\vec{m}}k_j^{\vec{m}}}{\avg{k^{\vec{m}}}N}\right]^{A_{ij}^{\vec{0}}}
\eea
where we have assumed that  the multidegrees $k_i^{\vec{m}}<\sqrt{\Avg{k_i^{\vec{m}}}N}$, $\forall \vec{m}\neq \vec{0}$, and we have indicated with  $\Avg{k_i^{\vec{m}}}$  the average of the multidegrees $\vec{m}$ present in the network. The condition on the multidegrees imposes  a ``multidegree structural cutoff" that ensure the fact that in the multiplex the multidegrees of linked nodes are not correlated. 
We assume that the nodes are removed with probability  $(1-p)$, and that therefore the sequence $\{s_i\}$ indicating if the node $i$ is removed ($s_i=0$) or not ($s_i=1$) has probability
\bea
P(\{s_i\})=\prod_{i=1}^N \left[p s_i+(1-p)(1-s_i)\right].
\eea 

We define   $S_{\vec{n}}$ the probability that following a multilink $\vec{n}$  (with $\sum_{\alpha}n_{\alpha}>0$) we reach a node in the MCGC in a multiplex in this ensemble. Since $\sigma_{j\to i}^{\vec{n}}=0,1$ the probability $S_{\vec{n}}$ that a random message in the multiplex is equal to one, i.e. $S_{\vec{n}}=P(\sigma_{j\to i}^{\vec{m}^{ij}}=1|\vec{m}^{ij}=\vec{n})$ is equal to the average  
\bea
S_{\vec{n}}&=&\overline{\Avg{\delta(\vec{m}^{ij},\vec{n})\sigma_{j\to i}^{\vec{m}^{ij}}}}
\label{snp1}
\eea
where $\delta(\vec{m}^{ij},\vec{n})=1$ if $\vec{m}^{ij}=\vec{n}$ and $\delta(\vec{m}^{ij},\vec{n})=0$ otherwise, and where we use the notation
\bea
&\overline{\Avg{\delta(\vec{m}^{ij},\vec{n})f_{ij}(\vec{m}^{ij})}}=\nonumber \\
&=\sum_{\vec{\G}}P(\vec{\G})\sum_{\{s_i\}}P(\{s_i\})\sum_{i,j}\frac{\delta(\vec{m}^{ij},\vec{n})f_{ij}({\vec{m}^{ij}})}{\sum_{i',j'}\delta(\vec{m}^{i'j'},\vec{n})}.
\eea 
for any function $f_{ij}(\vec{m}^{ij})$.

The equation for $S_{\vec{n}}$ can be derived from the recursive Eq.~$(\ref{sij})$ as follows
\bea
S^{\vec{n}}&=&\overline{\Avg{\sigma_{j\to i}^{\vec{m}^{ij}}\delta(\vec{m}^{ij},\vec{n})}}\label{snp}\\
&&\hspace*{-7mm}=\overline{\Avg{\delta(\vec{m}^{ij},\vec{n})\ s_j\sum_{\vec{r}}(-1)^{\sum_{\alpha}r_{\alpha}}\prod_{
		\begin{subarray}{c}
        			l=1\\
			l\neq i
      		\end{subarray}}^N (1-\sigma_{l \to j}^{\vec{m}^{lj}})^{\sum_{\alpha}r_{\alpha}m^{lj}_{\alpha}}}}.\nonumber
\eea
The probability that a random multilink $\vec{n}=\vec{m}^{ij}$  of the network will have a node at its end (node $j$) with  multidegree sequence $\{k^{\vec{m}}\}$ is given by
$k^{\vec{n}}P(\{k^{\vec{m}}\})/\Avg{k^{\vec{n}}}$.  If the node $j$ has multidegrees $k^{\vec{m}}$, it will have a number $k^{\vec{m}}$ of incoming messages from multilinks $\vec{m}$. When calculating the average in Eq.$(\ref{snp})$ we have to consider all the messages $\sigma_{l\to j}^{\vec{m}^{lj}}$ incoming to node $j$ except the message coming from node $i$ that is given by $\sigma_{i\to j}^{\vec{n}}$, \ie,  it is of type $\vec{n}$. Therefore, since  on a tree-like network the messages sent by any two neighbors will be independent,   we have that Eq. $(\ref{snp})$ becomes
\bea
S_{\vec{n}}&=&p \sum_{\{k^{\vec{m}}\}} \frac{k^{\vec{n}}}{\Avg{k^{\vec{n}}}}P(\{k^{\vec{m}}\})\sum_{\vec{r}}(-1)^{\sum_{\alpha}r_{\alpha}}\times\nonumber \\
&&\times \prod_{\vec{m}|\sum_{\alpha}m_{\alpha}r_{\alpha}>0} {\overline{\Avg{(1-\sigma_{l \to j}^{\vec{m}^{lj}})\delta(\vec{m},\vec{m}^{lj})}}}^{k^{\vec{m}}-\delta(\vec{m},\vec{n})}, \nonumber \\
\label{Snu}
\eea
where $P(\{k^{\vec{m}}\})$ is the multidegree distribution, the sum $\sum_{\vec{r}}$ is over all the possible vectors $\vec{r}=(r_1,r_2,\ldots, r_{\alpha},\ldots, r_M)$. 
Finally using the definition of $S_{\vec{m}}$ given in Eq. $(\ref{snp1})$ we obtain,
\bea
S_{\vec{n}}&=&p\sum_{\{k^{\vec{m}}\}}\frac{k^{\vec{n}}}{\Avg{k^{\vec{n}}}}P(\{k^{\vec{m}}\})\sum_{\vec{r}}(-1)^{\sum_{\alpha=1}^M r_{\alpha}}\times \nonumber \\
&&\times \prod_{\vec{m}|\sum_{\alpha}m_{\alpha}r_{\alpha}>0}(1-S_{\vec{m}})^{k^{\vec{m}}-\delta({\vec{m},\vec{n}})},
\label{int_g2}
\eea
where we have used the same notation as in  Eq. $(\ref{Snu})$.
 Similarly, the probability $S=P(\sigma_i=1)$ that a random node belongs to the MCGC can be expressed as
 \bea
S&=&\sum_{\vec{\G}}P(\vec{\G})\sum_{\{s_i\}}P(\{s_i\})\sum_{i}\frac{\sigma_{i}}{N}.
\eea
 Starting from Eq.~$(\ref{s2})$ and using similar steps used to derive Eq. $(\ref{int_g2})$, it can be shown that $S$
must satisfy the following equation
\begin{eqnarray}
	S&=&p\sum_{\{k^{\vec{m}}\}}P(\{k^{\vec{m}}\})\sum_{\vec{r}}(-1)^{\sum_{\alpha=1}^M r_{\alpha}} \times \nonumber \\
	&&\times \prod_{\vec{m}|\sum_{\alpha}m_{\alpha}r_{\alpha}>0}(1-S_{\vec{m}})^{k^{\vec{m}}},
	\label{int_g0}
\end{eqnarray}
where we have used the same notation as in  Eq. $(\ref{Snu})$.

We note here that the Eqs. $(\ref{int_g0})-(\ref{int_g2})$ are generalizations of Eq. $(8)$ in \cite{Havlin2b}, when the overlap of the links is significant. Moreover we observe that here, as in other percolation problems \cite{Kabashima}, the quantities $S_{\vec{n}}$ and $S$ are self-averaging, i.e., we expect that the network average of the messages and the values of $\sigma_i$ {converge,} in the large network limit,  to $S_{\vec{n}}$ and $S$ {respectively}.

An interesting case emerges when the degree distribution of the multidegrees
is factorizable, {\ie} $P(\{k^{\vec{m}}\})=\prod_{\vec{m}}p_{k^{\vec{m}}}$ where $p_{k^{\vec{m}}}$ is the degree distribution of the multidegree $k^{\vec{m}}$. In this case  we can introduce the following generating functions of the real (scalar) variable $z_{\vec{m}}$:
\begin{eqnarray}
	G^0_{\vec{m}}(z_{\vec{m}})&=&\sum_{k^{\vec{m}}} p_{k^{\vec{m}}} z_{\vec{m}}^{k^{\vec{m}}} \nonumber \\
	G^1_{\vec{m}}(z_{\vec{m}})&=&\sum_{k^{\vec{m}}}\frac{k^{\vec{m}}}{\Avg{k^{\vec{m}}}} p_{k^{\vec{m}}} z_{\vec{m}}^{k^{\vec{m}}-1}.
\end{eqnarray}
Therefore Eqs. (\ref{int_g0}) and (\ref{int_g2}) now read
\begin{equation}
	S = p\sum_{\{k^{\vec{m}}\}}\sum_{\vec{r}}(-1)^{\sum_{\alpha=1}^M r_{\alpha}}\prod_{
		\begin{subarray}{c}
        			\vec{m}\\
			\sum_{\alpha}m_{\alpha}r_{\alpha}>0
      		\end{subarray}
	}G^0_{\vec{m}}(1-S_{\vec{m}})
\end{equation}
\begin{eqnarray}
	S_{\vec{n}}&=&p\sum_{\{k^{\vec{m}}\}}\sum_{\vec{r}}(-1)^{\sum_{\alpha=1}^M r_{\alpha}}\left[G^1_{\vec{n}}(1-S_{\vec{n}})\right]^{f({\vec{n},\vec{r})}}\times \nonumber \\
	&&\times \prod_{
		\begin{subarray}{c}
        			\vec{m}\\
			\sum_{\alpha}m_{\alpha}r_{\alpha}>0\\
			\vec{m}\neq \vec{n}
      		\end{subarray}
	}G^0_{\vec{m}}(1-S_{\vec{m}})
\end{eqnarray}
where $f(\vec{n},\vec{r})=1$ if $\sum_{\alpha}r_{\alpha}n_{\alpha}>0$ and   $f(\vec{n},\vec{r})=0$ otherwise.

\section{Specific examples}

\subsection{ Two Poisson layers with Overlap}
We consider now the case of a duplex $M=2$ in which the multi-degree distributions are Poisson with means governed by the real parameters $c^1$ and $c^2$ (note that, here and in the following, $1$ and $2$ in $c^1$ and $c^2$ are indices, not exponents). 

\begin{figure}[htb]
\begin{center}
	\includegraphics[width=0.99\columnwidth]{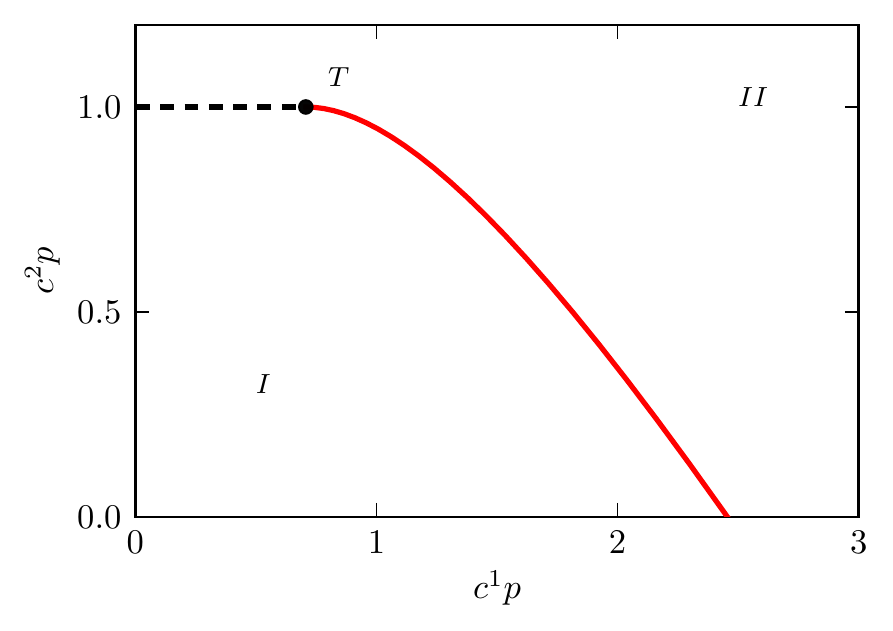}
	\caption{(Color online) Phase diagram of the two layer multiplex with Poisson degree distribution in each layer.
	In region I there is no percolation, in region II the system supports a MCGC.
	The solid red line indicates the points of hybrid first order phase transitions, the dashed black line indicates the line of second order  phase transitions. $T$ is the tricritical point. }
	\label{fig:Figure1}
\end{center}
\end{figure}

So we assume:  $\avg{k^{11}}=c^2$, and $\avg{k^{01}}=\avg{k^{10}}=c^1$.
Due to the properties of the Poisson distribution, from equation (\ref{int_g0}) we have
$S=S_{01}=S_{10}=S_{11}$, where $S$ satisfies the  equation
\begin{equation}
	S=p\left[1-2e^{-(c^1+c^2)S}+e^{-(2c^1+c^2)S}\right].
	\label{int_simple}
\end{equation}
By setting $x=S/p$, we can study the solutions of the equivalent equation
\begin{equation}
	h(x)=x-\left[1-2e^{-(c^1p+c^2p)x}+e^{-(2c^1p+c^2p)x}\right]=0
\end{equation}
in the  $(c^1p,c^2p)$ parameter plane.
Fig.~\ref{fig:Figure1} shows the phase diagram of this model.
Here the red solid line indicates a line of hybrid first order phase transition points and the black dashed line indicates a line of second order phase transition points, the point $T$ is a tricritical point.

\begin{figure}[htb]
\begin{center}
	\includegraphics[width=0.99\columnwidth]{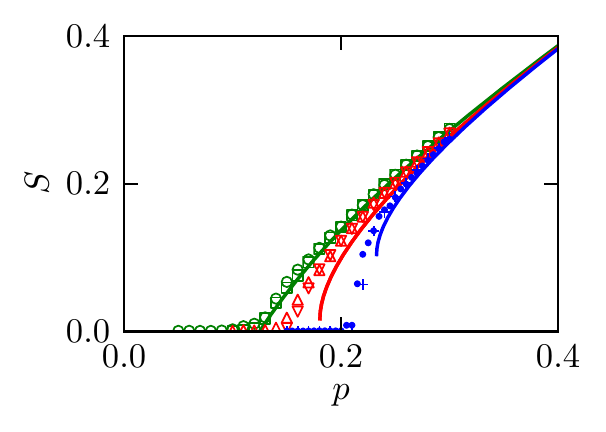}
	\caption{(Color online)
	Strength $S$ of the MCGC as a function of $p$ for some values of $c^2/c^1$ (here we only show a zoom of the region where the transitions occur).
	Continuous lines represent the analytic solution, points indicates simulations over networks of size $N$.
	From left to right: $c^2/c^1 =  4$ (green),  $N=10^4$ ($\bigcirc$), $N=5\cdot 10^4$ ($\square$); $c^2/c^1 =  1.22$ (red),  $N=10^4$ ($\bigtriangleup$), $N=5\cdot 10^4$ ($\bigtriangledown$), $c^2/c^1 =  0.25$ (blue), $N=10^4$ ($\bullet$), $N=5\cdot 10^4$ ($+$).
	From the comparison between the two sizes, it appears that the points converge to the analytical solution when $N$ increases. 
	}
	\label{fig:2er-simulations}
\end{center}
\end{figure}

We compare the analytical solutions $S(p)$ with numerical simulations in Fig.~\ref{fig:2er-simulations}.
There is good agreement where the transition is continuous.
In the case of discontinuous transitions and close to the tricritical point, we observe finite size effects, with an improved  agreement  for larger network sizes.

The line of continuous phase transitions  can be calculated analytically by imposing the condition that a non trivial solution $x^{\star}>0$, satisfying $h(x^{\star})=0$, goes to zero as a function of the parameter $p$.
Therefore, we expand $h(x^{\star})$ for $x^{\star}=\epsilon \ll 1$ finding
\begin{equation}
	 h(x^{\star})=h^{\prime}(0)\epsilon+\frac{1}{2} h^{\prime \prime}(0)\epsilon^2+\frac{1}{3!} h^{\prime\prime\prime}(0)\epsilon^3 +{\cal O}(\epsilon^4).
	 \label{h-expansion}
\end{equation}
If $h^{\prime}(0)<0$ and $h^{\prime\prime}(0)>0$, we find the following  solution $x^{\star}=\epsilon\ll1$ of Eq.~$(\ref{h-expansion})$
\begin{equation}
	x^{\star}=\epsilon\propto \left(p-1/c^2\right),
\end{equation}
implying
\begin{equation}
	S\propto \left(p-1/c^2\right)^{\beta}
\end{equation}
with $\beta=1$.
This indicates that as long as $h^{\prime \prime}(0)>0$ ({\ie} $c^2>\sqrt{2}c^1$) the points $c^2p=1$ for which $h^{\prime}(0)=0$ are second order critical points.

At the point $c^2/c^1=\sqrt{2}$, $c^2p=1$, we have  $h^{\prime\prime}(0)=0$.
To find a non trivial solution of Eq.~$(\ref{h-expansion})$ for $p\simeq p_c=1/c^2$ (in the following, we will use $p_c$ as the generic critical value of $p$ at a phase transition), we have to go up to the third order in the $\epsilon$ expansion, finding,  since $h^{\prime\prime\prime}(0)>0$,
\begin{equation}
	x^{\star}=\epsilon\propto \left(p-1/c^2\right)^{1/2},
\end{equation}
implying
\begin{equation}
	S\propto \left(p-1/c^2\right)^{\beta}
\end{equation}
with $\beta=1/2$. Therefore the  point $c^2/c^1=\sqrt{2}$, $c^2p=1$ is the tricritical point $T$.
For $c^2<\sqrt{2}c^1$ and $c^2p<1$ we observe a line of first order phase transition points determined by the conditions $h({x^{\star}})=h^{\prime}(x^{\star})=0$ with $x^{\star}>0$.
The expansion of Eq.~(\ref{int_simple}) at these transition points shows that $\beta=1/2$, thus the transition is hybrid.
Finally, we observe that in the case $c^2=0$ we have $c^1 p_c =2.4554\ldots$ recovering the result of two Poisson networks  without overlap and with average degree $c^1$ \cite{Havlin1}.


From this simple model we observe several interesting features.
First, we observe that increasing the overlap  $c^2$ between the layers improves the robustness of the multiplex system and changes the order of the percolation phase transition from first order to second order in a smooth way (through a tricritical point).
Moreover, the continuous phase transition is entirely driven by the (classical) percolation of the sub-network of double edges ($c^2p=1$).
Finally, it is interesting to note that when the ratio $c^1/c^2$ is large enough, the percolating phase extends in a region where it can be $c^2 p <1$ (Fig.~\ref{fig:Figure1}).
This is not surprising, as it signals that the small overlap behavior takes over the overlap-driven percolating phase.
In other words, on the right hand side of the tricritical point $T$ in Fig.~\ref{fig:Figure1}, the MCGC is not simply containing a classical giant cluster entirely constituted by double edges, as this does not percolate on its own if $c^2 p <1$.
Instead, a relevant fraction of pairs of nodes in the MCGC are connected by non-coincident paths of each type of edges.
This is the scenario described, in the case of negligible edge overlap, in \cite{baxter2012}.

\subsection{ Three  Poisson Layers with Overlap}
As a second example, we consider the ensemble of a three layer multiplex $(M=3)$ with Poisson multi-degree distribution and $\avg{k^{100}}=\avg{k^{001}}=\avg{k^{010}}=c^1$, $\avg{k^{110}}=\avg{k^{101}}=\avg{k^{011}}=c^2$ and $\avg{k^{111}}=c^3$ (recall that 1, 2, 3 in $c^1$, $c^2$, $c^3$ are, as earlier, indices, not exponents).
As in the previous case, we have just one order parameter $S=S_{\vec{m}}$ that satisfies the equation
\begin{eqnarray}
	S &=& p \left[1-3e^{-(c^1+2c^2+c^3)S} \right.\nonumber\\
	&&+ \left. 3e^{-(2c^1+3c^2+c^3)S}-e^{-(3c^1+3c^2+c^3)S}\right].
	\label{3er-S-equation}
\end{eqnarray}
By setting  $x=S/p$, we can define a function $g(x)$ as
\begin{eqnarray}
	g(x)&=&x-[1-3e^{-(c^1p+2c^2p+c^3p)x}+3e^{-(2c^1p+3c^2p+c^3p)x}\nonumber \\
	&& -e^{-(3c^1p+3c^2p+c^3p)x}],
	\label{gx}
\end{eqnarray}
and we can recast the equation for $S$ as $g(x^{\star})=0$.

\begin{figure}[htb]
\centering
\includegraphics[width=0.4\textwidth]{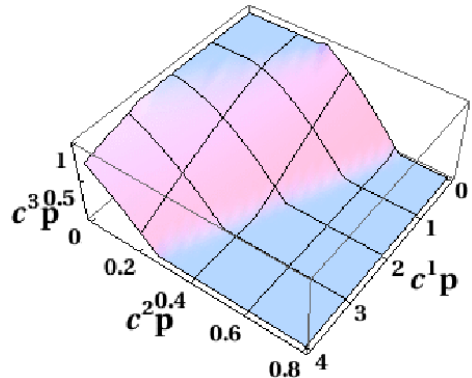}
\caption{(Color online) The phase diagram of the three layer multiplex with Poisson multidegree distribution parametrized by $c^1,c^2,c^3$.
Below the plotted surface we have $S=0$, above the surface we have $S>0$.}
\label{fig:3D}
\end{figure}

The 3D phase diagram is displayed in Fig.~\ref{fig:3D} for $c^3p<1$.
Under the surface there is no percolation ($S=0$), whereas above the surface we have $S>0$.
The blue surface at $c^3 p = 1$ corresponds to continuous transitions, the pink surface at $c^3p<1$ represents discontinuous transitions.

\begin{figure}[htb]
	\includegraphics[width=0.99\columnwidth]{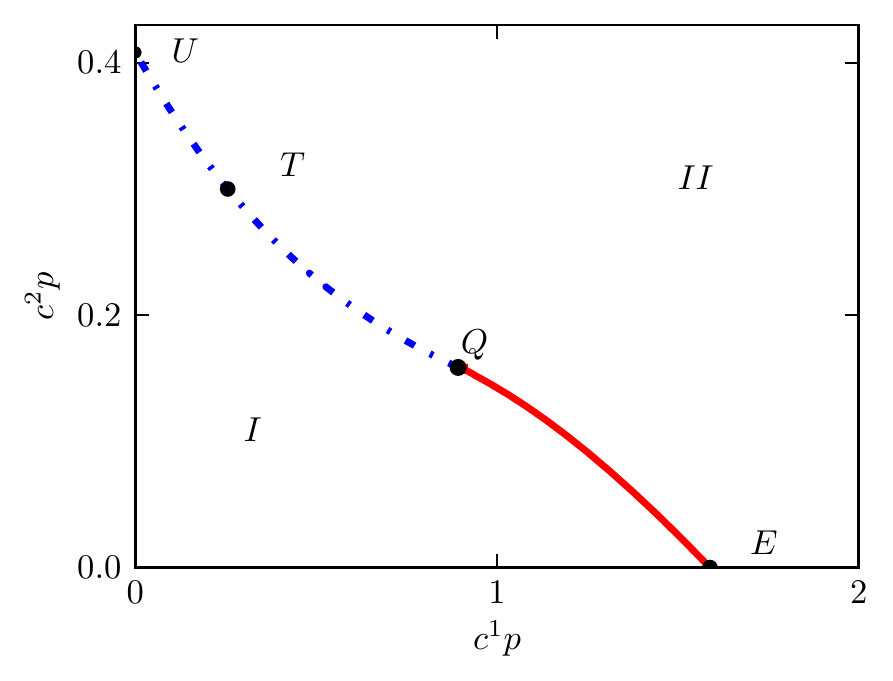}
	\caption{(Color online) Section of the phase diagram of the three layer multiplex with Poisson multidegree distribution at $c^3p=1$.
	A line of tricritical points (dot-dashed blue) encounters the line of discontinuous phase transitions (solid red) at the multicritical point $Q$.
	}
	\label{fig:er3-phase-c1c2}
\end{figure}

Fig.~\ref{fig:er3-phase-c1c2} displays a section of the same phase diagram at $c^3 p =1$.
This section
is quite relevant, as it entirely contains the line of tricritical points, plotted as a dot-dashed blue line.
This line starts at point  $U$ given by $c^1p=0$, $c^2p=1/\sqrt{6},c^3p=1$ and terminates at a multicritical  point $Q$ determined  by the conditions $g(0)=g^{\prime}(0)=g^{\prime\prime}(0)=g^{\prime\prime\prime}(0)=0$ given by
$c_Q^1p=0.892550$, $c_Q^2p=0.158562$, $c_Q^3p=1$ (see Appendix for further details).
The tricritical line is determined by the condition $g(0)=g^{\prime}(0)=g^{\prime\prime}(0)=0$ and is given by
 $c^2p=\frac{1}{6}[-3c^1p+\sqrt{6+9(c^1p)^2}],c^3p=1$, with $c^1p<c_Q^1p$.
At $c^1p>c_Q^1p$, the nature of the transition changes, as the relatively higher fraction of single edges drives the system into a behavior where the MCGC is mainly characterized by nodes connected by different paths on each layer (as in the case of non overlapping layers \cite{baxter2012}).
Therefore, the solid red line beyond point $Q$ represents the locus of points  characterized by a discontinuous phase transition of the percolating phase into a critical $S=0$ phase.

\begin{figure}[htb]
	\includegraphics[width=0.99\columnwidth]{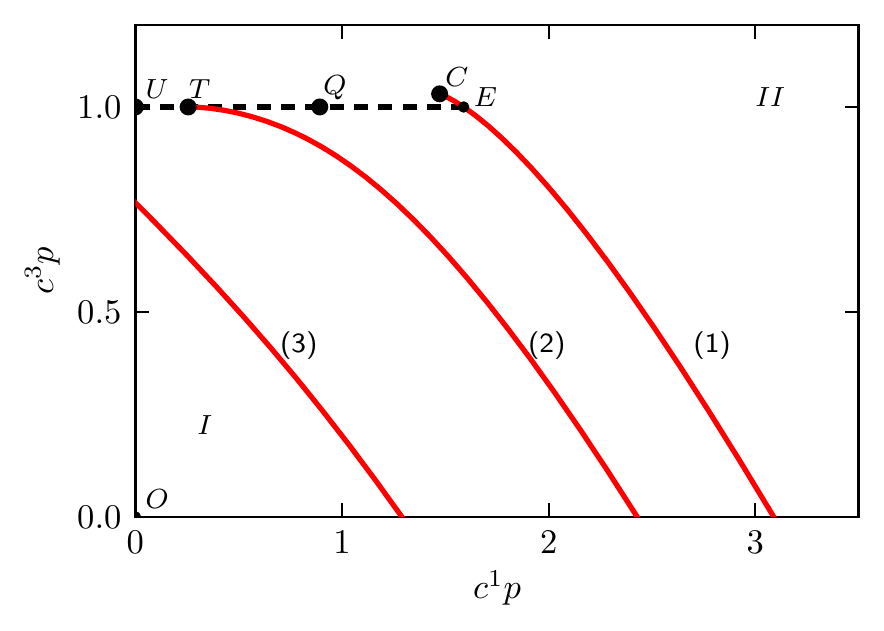}	
	\caption{(Color online) Phase diagram of the three layer multiplex with Poisson multidegree distribution at fixed $c^2 p$.
		The dashed black line is the line of second order transitions (classical percolation).
		The red full lines represent first order hybrid transitions at different values of $c^2 p$.
		Lines (1-3) refer to $c^2 p = 0$, $0.3$ and $0.8$, respectively.
	}
	\label{fig:5}
\end{figure}

To further characterize these critical phenomena, it is also interesting to examine sections at fixed $c^2 p$ (Fig.~
\ref{fig:5}), which essentially governs the fraction of double edges in the multiplex network.
We can identify three characteristic topologies as $c^2p$ varies, exemplified by the lines (1), (2) and (3) of Fig.~\ref{fig:5}.
The two discriminating cases are characterized by (i) the multicritical point $Q$, and (ii) the end point $U$.

At small values of $c^2p$, the MCGC can collapse either continuously or discontinuously, depending on the relative abundance of single and triple edges.
If $c^3/c^1$ is smaller than the slope of the segment $OE$, random damage causes a discontinuous transition, whereas along the $c^1p=0$ line, for example, we recover the continuous phase transition of classical percolation as the three Poisson networks are coincident.
The line  of second order critical points (dashed black) is determined by the condition $g(0)=g^{\prime}(0)=0$,  which implies $c^3p=1$.
Similarly to the two layer case, the continuous transition constitutes the classical percolation scenario and is entirely driven by the subnetwork of triple edges.
This line ends as it encounters the line of discontinuous transitions at a point which depends on the value of $c^2p$.
Quite interestingly, the line of first order hybrid transitions extends into the region $c^3p>1$ and ends in a critical point $C$ defined by $g(x^{\star})=g^{\prime}(x^{\star})=g^{\prime\prime}(x^{\star})=0$ with $x^{\star}> 0$.
Indeed there is a region of phase space such that if we fix $c^1$, $c^2$ and $c^3$ and we  raise the value of $p$, we first cross a second order transition point as $S$ continuously grows to a small finite value,  then we cross an additional discontinuous phase transition and $S$ jumps from a small finite value to a larger value.

\begin{figure}[ht]
		\includegraphics[width=0.9\columnwidth,angle=0]{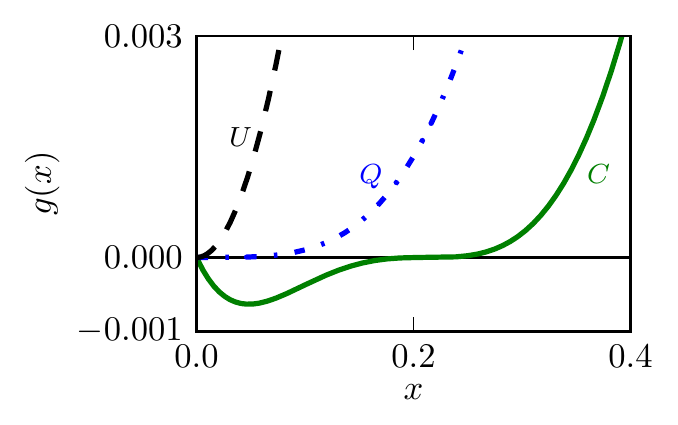}
		\caption{(Color online) Behavior of the function $g(x)$ at the  points $U$, $Q$, $C$ and $E$. The point $Q$ is characterized by the conditions $g(0)=g^{\prime}(0)=g^{\prime\prime}(0)=g^{\prime\prime\prime}(0)=0$. The points $C$ are characterized by the conditions $g(x^{\star})=g^{\prime}(x^{\star})=g^{\prime\prime}(x^{\star})=0$ with $x^{\star}>0$.
}
	\label{fig:6}
\end{figure}

This discontinuous transition separates a percolating phase (driven by nodes connected by non-coincident paths) from a classical percolating phase (driven by coincident paths formed by triple edges).
In the second characteristic topology, occurring for $c_U^2 < c^2< c_Q^2$, the critical line and the line of discontinuous transitions match at a tricritical point $T$.
The line of tricritical points ends at the multicritical point $Q$.
The second order critical line, instead, disappears at the point $U$.
For $c^2p>1/\sqrt{6}$, the third topology is only characterized by a line of discontinuous transitions driven by the low overlap behavior (line (3)).
In Fig. $\ref{fig:6}$, we plot the behavior of the function $g(x)$ at the points $U$, $Q$ and one of the critical points $C$ of the phase diagram.

As is customary in the analysis of critical phenomena, the exponent $\beta$ determines the critical behavior of $S$ in the vicinity of a phase transition occurring at a critical occupation probability $p_c$: $S-S_c\propto (p-p_c)^{\beta}$ as $p\to p_c$.
The values of $\beta$ can be calculated at every critical point by an appropriate expansion of the equation (\ref{3er-S-equation}), as explained in detail in the Appendix.
We find $\beta=1$  at the  second order transition points, $\beta=1/2$ at the  tricritical points, $\beta=1/3$ at the multicritical point $Q$ and $\beta=1/3$ at the critical points $C$.
This implies that the critical points $C$ are in the same universality class as the critical points observed in heterogeneous $k$-core percolation on Poisson networks \cite{baxter2011, cellai2013a}.

As in the two layer case, this phase diagram shows that  the multiplex becomes more resilient by increasing the fraction of triple edges.
However, it also shows that by raising the ratio  $c^2/c^1$ the phase transition may become discontinuous.
For example, if we consider Fig.~\ref{fig:5}, it emerges that at fixed $c^1 p$ one may encounter a discontinuous transition by varying the parameter $c^3 p$.
This occurs if the relative fraction of double edges $c^2 p$ is large enough.
This suggests that network design must take into consideration \emph{all} the  relevant layers, otherwise network failure might evolve towards a catastrophic regime.

\section{ Conclusions}
In this paper we have presented a general framework for studying the emergence of a MCGC in multiplex network with overlap.
We show that the presence of a critical value of edge overlap in a duplex can both change the order of the phase transition from hybrid first order to second order, and improve the robustness of the system.
We also show that in multiplexes with more than two layers the observed critical phenomena	become remarkably complex, including the presence of high order multicritical points.
On one hand, there may occur first order phase transitions between percolating phases with different strengths.
On the other, it emerges that overlap of edges which do not involve all the layers can also change a continuous phase transition to a discontinuous one, making system reliability less predictable.
This is a feature which may be relevant in the design of large scale infrastructures.
Here we stress that in real multiplexes---such as in online games \cite{Thurner}, social networks \cite{cozzo2013} and epidemiology \cite{funk2010}---the presence of link overlap is  the norm rather than the exception.
Therefore,  this work represents an important step for characterizing the robustness properties of real multiplexes, and it is also likely to have an impact on the dynamic processes occurring on multiplex systems.

Note: During the evaluation of this manuscript, we came to know about two papers where percolation on 2-layer Poisson graphs with overlap is studied \cite{li2013, hu2013}.

D.~C., J.~Z.~and J.~G.~ acknowledge funding from Science Foundation Ireland (11/PI/1026) and the FET-Proactive project PLEXMATH and E.~L.~acknowledges funding from James Martin 21st Century Foundation Reference no: LC1213-006.

\appendix

\section{Calculation of the critical points in the three layer Poisson multiplex}


We now calculate the critical points and exponents of the 3-layer Poisson multiplex with $\avg{k^{100}}=\avg{k^{001}}=\avg{k^{010}}=c^1$, $\avg{k^{110}}=\avg{k^{101}}=\avg{k^{011}}=c^2$ and $\avg{k^{111}}=c^3$.
Let $x^{\star}>0$ be a solution of equation $g(x)=0$ where the function $g$ is defined in (\ref{gx}).
In order to calculate the position of a continuous phase transition we expand $g(x^{\star})$ for $x^{\star}=\epsilon\ll 1$ finding
\begin{eqnarray}
	 g(x^{\star})&=&g^{\prime}(0)\epsilon+\frac{1}{2} g^{\prime \prime}(0)\epsilon^2+\nonumber \\
	 &&+\frac{1}{3!} g^{\prime\prime\prime}(0)\epsilon^3+\frac{1}{4!}g^{\prime\prime\prime\prime}(0)\epsilon^4 +{\cal O}(\epsilon^5),
	 \label{d2}
\end{eqnarray}
with
\begin{widetext}
\begin{eqnarray}
	g^{\prime}(0)&=&1-c^3p\nonumber \\
	g^{\prime \prime}(0)&=&3(c^1p+c^2p+c^3p)^2-3(2c^1p+3c^2p+c^3p)^2-(3c^1p+3c^2p+c^3p)^2\nonumber \\
	g^{\prime \prime \prime}(0)&=& -3(c^1p+c^2p+c^3p)^3+3(2c^1p+3c^2p+c^3p)^3-(3c^1p+3c^2p+c^3p)^3\nonumber \\
	g^{\prime\prime\prime\prime}(0)&=&3(c^1p+c^2p+c^3p)^4-3(2c^1p+3c^2p+c^3p)^4-(3c^1p+3c^2p+c^3p)^4.
\end{eqnarray}
\end{widetext}
The set of second order critical points are determined by the condition $g^{\prime}(0)=0$ and $g^{\prime\prime}(0)>0$ with the additional condition that a first order phase transition has not already occurred in the multiplex.
These conditions imply $c^3p=1$ and determine all the points in region $I$ of Fig.~\ref{fig:5}  in the main text.
Close to these transition points, for $p\simeq p_c=1/c^3$, we have $g^{\prime\prime}(0)>0$ therefore, by using the expansion of $g(x)$ given by Eq. $(\ref{d2})$ we get
\begin{equation}
	x^{\star}=\epsilon\propto \left(p-1/c^3\right),
\end{equation}
implying
\begin{equation}
	S\propto \left(p-1/c^3\right)^{\beta}
\end{equation}
with $\beta=1$.

The line of tricritical points is determined by the conditions
$g^{\prime}(0)=g^{\prime\prime}(0)=0$ and $g^{\prime\prime\prime}(0)>0$ yielding
\begin{eqnarray}
	c^1p&<&\frac{3+\left(3-2\sqrt{2}\right)^{1/3}+\left(3+2\sqrt{2}\right)^{1/3}}{6} \nonumber \\
	c^2p&=&\frac{1}{6}\left[-3c^1p+\sqrt{6+9 (c^1p)^2}\right] \\
	c^3p&=&1.\nonumber
\end{eqnarray}
Close to these transition points, for $p\simeq p_c=1/c^3$, we have $g^{\prime\prime\prime}(0)>0$ therefore, by using the expansion of $g(x)$ given by Eq. $(\ref{d2})$ we get
\begin{equation}
	x^{\star}=\epsilon\propto \left(p-1/c^3\right)^{1/2},
\end{equation}
implying
\begin{equation}
	S\propto \left(p-1/c^3\right)^{\beta}
\end{equation}
with $\beta=1/2$.
The multicritical point $Q$ is determined by the conditions $g^{\prime}(0)=g^{\prime\prime}(0)=g^{\prime\prime\prime}(0)=0$ yielding
\begin{eqnarray}
	\frac{c^1}{c^3}&=&\frac{3+\left(3-2\sqrt{2}\right)^{1/3}+\left(3+2\sqrt{2}\right)^{1/3}}{6}=0.89255...\nonumber \\
	\frac{c^2}{c^3}&=&\frac{-2+\left(4-2\sqrt{2}\right)^{1/3}+\left(4+2\sqrt{2}\right)^{1/3}}{6}=0.158562....\nonumber \\
	p&=&1/c^3.
\end{eqnarray}
Close to this transition point, for $p\simeq p_c=1/c^3$, we have $g^{\prime\prime\prime\prime}(0)>0$.
Therefore, by using the expansion of $g(x)$ given by Eq. $(\ref{d2})$, we get
\begin{equation}
	x^{\star}=\epsilon\propto \left(p-1/c^3\right)^{1/3},
\end{equation}
implying
\begin{equation}
	S\propto \left(p-1/c^3\right)^{\beta}
\end{equation}
with $\beta=1/3$.

Regarding the critical points $C$ occurring at $x_c>0$, we have the conditions $g(x_c)=g'(x_c)=g''(x_c)=0$.
Let us define the following auxiliary function $\Phi(y^1,y^2,y^3)$ (as throughout the paper, $\alpha$ is a layer index, not an exponent, in variable $y^{\alpha}$):
\begin{eqnarray}
	\Phi(y^1,y^2,y^3) &=& 1-3e^{-(y^1+2y^2+y^3)}+3e^{-(2y^1+3y^2+y^3)} \nonumber \\
	&&-e^{-(3y^1+3y^2+y^3)}.
\end{eqnarray}
Using function  $\Phi$, we can rewrite $g(x)$ and its derivatives in { the} following way:
\begin{equation}
	g(x) = x - \Phi(c^1px,c^2px,c^3px) = \frac{S}{p} - \Phi(c^1S,c^2S,c^3S)
	\label{eq:gx-C}
\end{equation}
\begin{equation}
	g'(x) = 1 - p\sum_i c^i \frac{\partial\Phi}{\partial y^i}
\end{equation}
\begin{equation}
	g''(x) = - p^2\sum_{ij} c^i c^j \frac{\partial^2\Phi}{\partial y^i\partial y^j}.
\end{equation}
So, the conditions $g(x_c)=g'(x_c)=g''(x_c)=0$ defining a point $C$ imply:
\begin{equation}
	 \frac{S_c}{p_c} = \Phi(c^1S_c,c^2S_c,c^3S_c)
\end{equation}
\begin{equation}
	\sum_i c^i \frac{\partial\Phi}{\partial y^i} = \frac{1}{p_c}
\end{equation}
\begin{equation}
	\sum_{ij} c^i c^j \frac{\partial^2\Phi}{\partial y^i \partial y^j} = 0
\end{equation}
Now let us expand equation (\ref{eq:gx-C}) by imposing $S = S_c + \xi$ for $\xi\to 0$ and $p = p_c +\delta$ for $\delta\to 0$.
Substituting the expansion of $\Phi$
\begin{widetext}
\begin{multline}
	\Phi(c^1(S_c+\xi),c^2(S_c+\xi),c^3(S_c+\xi)) = \Phi(c^1S_c,c^2S_c,c^3S_c) + \sum_i c^i \frac{\partial\Phi}{\partial y^i} \xi \\
	+ \frac{1}{2} \sum_{ij} c^i c^j \frac{\partial^2\Phi}{\partial y^i \partial y^j} \xi^2
	+ \frac{1}{6} \sum_{ijk} c^i c^j  c^k \frac{\partial^3\Phi}{\partial y^i \partial y^j \partial y^k} \xi^3 + O(\xi^4),
\end{multline}
\end{widetext}
and applying the conditions above, we get
\begin{equation}
	0 = S_c + \xi- (p_c +\delta) \left[ \frac{S_c}{p_c} + \frac{\xi}{p_c} + K_3 \xi^3 + O(\xi^4)\right].
\end{equation}
 From which it yields
\begin{equation}
	\xi \sim \delta^{1/3}.
\end{equation}
Hence, we have $\beta=1/3$.


\begin{thebibliography}{99}


\bibitem{RMP} R. Albert and A.-L. Barabasi,
   {Reviews of Modern Physics} \textbf{74}, 47 (2002).

\bibitem{Newman_rev} M. E. J. Newman,  {SIAM Review}
  \textbf{45}, 167 (2003).

\bibitem{Boccaletti2006} S. Boccaletti, V. Latora, Y. Moreno,
  M. Chavez and D.-U. Hwang,  {Physics Reports} \textbf{424},
  175  (2006).


\bibitem{crit} S. N. Dorogovtsev, A. Goltsev and J. F. F. Mendes, Rev. Mod. Phys. {\bf 80}, 1275 (2008).

\bibitem{Dynamics}
A. Barrat, M. Barth\'elemy, A. Vespignani {\it Dynamical Processes on complex Networks} (Cambridge University Press, Cambridge, 2008).

\bibitem{Havlin1}
S. V. Buldyrev, R. Parshani, G. Paul, H. E. Stanley and S. Havlin,
Nature {\bf 464}, 1025 (2010).

\bibitem{Rinaldi} S. M. Rinaldi, J. P.  Peerenboom, and T. K. Kelly, IEEE Control Syst. Mag. {\bf 21}, 11 (2001).
\bibitem{Kurant}
M. Kurant and P. Thiran, Phys. Rev. Lett. {\bf 96}, 138701 (2006).

\bibitem{rosato2008} V. Rosato, L. Issacharoff, F. Tiriticco, S. Meloni, S. D. Porcellinis, and R. Setola, International Journal of Critical Infrastructures \textbf{4}, 63 (2008).


\bibitem{morris2012}
R. G. Morris, M. Barth\'elemy, Phys. Rev. Lett. {\bf 109}, 128703 (2012).


\bibitem{Boccaletti}
A. Cardillo, J. G\'omez-Garde\~nes, M. Zanin, M. Romance, D. Papo, F. del Pozo and S. Boccaletti, Sci. Rep. {\bf 3}, 1344 (2013).

\bibitem{Thurner}
M. Szell, R. Lambiotte, S. Thurner, PNAS, {\bf 107}, 13636 (2010).

\bibitem{Kurths}
J. Doges, H. Schultz, N. Marwan, Y. Zou and  J. Kurths, Eur. Phys. J. B {\bf 84}, 635 (2011).



\bibitem{Bullmore2009} E. Bullmore and O. Sporns,  {Nat Rev  Neurosci} \textbf{10}, 186 (2009).

  \bibitem{Mucha}
 P. J. Mucha, T. Richardson, K. Macon, M. A Porter, J.-P. Onnela,
Science, {\bf 328},876 (2010).

\bibitem{duenas2012} L. Due\~{n}as-Osorio and A. Kwasinski, Earthquake Spectra \textbf{28}, S581 (2012).

\bibitem{hernandez2011} I. Hernandez-Fajardo and L. Due\~{n}as-Osorio, Earthquake Spectra  \textbf{27}, 23 (2011).


\bibitem{MultiplexPR} A. Halu,  R. J. Mondragon, P. Panzarasa and G. Bianconi, PLoS ONE 8(10): e78293 (2013).

\bibitem{woolley2011}
O. Woolley-Meza, C. Thiemann, D. Grady, J. J. Lee, H. Seebens, B. Blasius, and D. Brockmann, The European Physical Journal B - Condensed Matter and Complex Systems 84, 589 (2011).

\bibitem{Growth1}
V. Nicosia, G. Bianconi, V. Latora, and M. Barthelemy, Phys. Rev. Lett. {\bf 111}, 058701 (2013).
\bibitem{Growth2}
J. Y. Kim, K.-I. Goh, Phys. Rev. Lett. {\bf 111}, 058702 (2013).
\bibitem{PRE}
G. Bianconi, Phys. Rev. E {\bf 87}, 062806 (2013).

\bibitem{dedomenico2013} M. De Domenico, et al., arXiv:1306.0519  (2013).

\bibitem{Havlin2}
R. Parshani, S. V. Buldyrev and S. Havlin, Phys. Rev. Lett. {\bf 105}, 048701 (2010).
\bibitem{Havlin2b}
J. Gao, S. V. Buldyrev, S. Havlin, H. E. Stanley,  Phys. Rev. Lett. {\bf 107}, 195701 (2011).
\bibitem{Havlin3}
J. Gao, S.V. Buldyrev, H.E. Stanley, S. Havlin, Nature Physics {\bf 8}, 40 (2012).

\bibitem{parshani2010}
R. Parshani, C. Rozenblat, D. Ietri, C. Ducruet, and S. Havlin, EPL (Europhysics Letters) 92, 68002 (2010).
\bibitem{baxter2011}
G. J. Baxter, S. N. Dorogovtsev, A. V. Goltsev and J. F. F. Mendes, Phys. Rev. E {\bf 83}, 051134 (2011).

\bibitem{cellai2013a}
D. Cellai, A. Lawlor, K. A. Dawson, and J. P. Gleeson, Physical Review E {\bf 87},  022134 (2013).

\bibitem{baxter2012}
G. J. Baxter, S. N. Dorogovtsev, A. V. Goltsev and J. F. F. Mendes, Phys. Rev. Lett. {\bf 109}, 248701 (2012).


\bibitem{Son}
S.-W. Son, G. Bizhani, C. Christensen, P. Grassberger and M. Paczuski, EPL {\bf 97} 16006 (2012).

\bibitem{Kabashima}
S. Watanabe and Y. Kabashima, ArXiv:1308.1210 (2013).
\bibitem{bollobas2001}
B{\'e}la Bollob{\'a}s, Random graphs  {\bf 73}, Cambridge university press (2001).

\bibitem{Leicht}
C. D. Brummitt, R. M. D'Souza, and E.A. Leicht,
PNAS {\bf 109}, 12 E680.
\bibitem{JSTAT}
K. Zhao and G. Bianconi,  J. Stat. Mech.  P05005 (2013).

\bibitem{Diffusion} S. G\'omez, A. D\'iaz-Guilera,
  J. G\'omez-Garde\~nes, C. J. P\'erez-Vicente, Y. Moreno and
  A. Arenas,  {Phys. Rev. Lett.} \textbf{110}, 028701 (2013).

\bibitem{funk2010}
S. Funk and V. A. A. Jansen, Physical Review E 81, 036118 (2010).
\bibitem{Boguna} A. Saumell-Mendiola, M. \'A. Serrano
  and M. Bogu\~n\'a,  {Phys. Rev. E} \textbf{86}, 026106 (2012).

\bibitem{Cooperation} J. Gomez-Garde\~nes, I. Reinares,
  A. Arenas and L. M. Floria,  {Sci. Rep.} \textbf{2}, 620 (2012).

\bibitem{cozzo2013} E. Cozzo, R. A. Ba\~{n}os, S. Meloni and Y. Moreno, Phys. Rev. E {\bf 88}, 050801 (2013).


\bibitem{Mezard}
M. Mezard and A. Montanari, {\it Information, physics and computation} (Oxford University Press, Oxford,2009).



\bibitem{li2013}
M. Li, R.-R. Liu, C.-X. Jia, and B.-H. Wang, New Journal of Physics {\bf 15}, 093013 (2013).

\bibitem{hu2013}
Y. Hu, D. Zhou, R. Zhang, Z. Han, and S. Havlin, Phys. Rev. E   {\bf 88}, 052805 (2013).


\end{thebibliography}
\end{document}